\documentclass[prl,a4paper,twocolumn,superscriptaddress,longbibliography]{revtex4-1}

\usepackage{amssymb}
\usepackage{amsmath}
\usepackage{amsfonts}
\usepackage{graphicx}
\usepackage{bm}
\usepackage{xcolor}
\usepackage{multirow}
\usepackage{natbib}
\usepackage{hyperref}
\usepackage[normalem]{ulem}
\usepackage{relsize}
\usepackage[final]{pdfpages}
\usepackage{pgffor}	
\usepackage{xspace}
\newcommand{\rrangle}{\rangle\kern-1.2ex~\rangle\xspace}
\newcommand{\llangle}{\langle\kern-1.2ex~\langle\xspace}

\usepackage{adjustbox}

\DeclareMathOperator{\Tr}{Tr}

\makeatletter
\AtBeginDocument{\let\LS@rot\@undefined}
\makeatother

\begin{document}

\title{Unrestricted electron bunching at the helical edge}

\author{Pavel D.~Kurilovich}

\affiliation{Department of Physics, Yale University, New Haven, CT 06520, USA}

\author{Vladislav D.~Kurilovich}

\affiliation{Department of Physics, Yale University, New Haven, CT 06520, USA}

\author{Igor S.~Burmistrov}

\affiliation{\hbox{L.~D.~Landau Institute for Theoretical Physics, acad. Semenova av. 1-a, 142432 Chernogolovka, Russia}}
  
  

\author{Yuval Gefen}

\affiliation{Department of Condensed Matter Physics, The Weizmann Institute of Science, Rehovot 76100, Israel}

\author{Moshe Goldstein}

\affiliation{Raymond and Beverly Sackler School of Physics and Astronomy, Tel Aviv University, Tel Aviv 6997801, Israel}

\begin{abstract}

A quantum magnetic impurity of spin $S$ at the edge of a two-dimensional time reversal invariant topological insulator may give rise to backscattering. We study here the shot noise associated with the backscattering current for arbitrary $S$.  Our full analytical solution reveals that for $S>\frac{1}{2}$ the Fano factor may be arbitrarily large, reflecting bunching of large batches of electrons. By contrast, we rigorously prove that for $S=\frac{1}{2}$ the Fano factor is bounded between $1$ and $2$, generalizing earlier studies. 
\end{abstract}


\maketitle

\textsf{Introduction.}\ ---  Zero-frequency current noise in a conductor can reveal information about correlations in  electronic transport which cannot be extracted from the average current \cite{BB,LS}. Obtaining information about the correlations  requires going beyond linear response (where thermal noise is fully determined by linear conductance through the fluctuation-dissipation theorem), and {studying} \emph{shot noise} at  voltage larger than the temperature. The ratio between the shot noise and the average current times the electron charge is referred to as \emph{Fano factor}. It is useful for characterizing  the unit of effective elementary charge in correlated electron systems, e.g., quasiparticle charges in fractional quantum Hall edges~\cite{Glattli,Reznikov}.  Entanglement with an external degree of freedom may modify the effective  Fano factor~\cite{Birchall,Barbarino}. 
\color{black} 

The experimental discovery of 2D topological insulators \cite{Konig2007} triggered intensive experimental and theoretical research \cite{Qi-Zhang,Hasan-Kane}. Electron transport along the helical edge was theoretically predicted to be protected from elastic backscattering by time-reversal symmetry. However, this ideal picture was impugned
by transport 
experiments in HgTe/CdTe ~\cite{Konig2007,Nowack,Grabecki,Gusev2013,Gusev2014,Kvon2015} and InAs/GaSb ~\cite{Knez2011,Suzuki2013,RRDu1,RRDu2,Suzuki2015,Mueller2015,RRDu3,Mueller2017} quantum wells, Bi bilayers~\cite{Sabater2013}, and WTe$_2$ monolayers~\cite{Cobden17,Jia2017,Herrero2018}. In order to explain this 
data, several physical mechanisms of backscattering were proposed and studied theoretically~\cite{Xu2006,Maciejko2009,Tanaka2011,Schmidt2012,Oreg2012,Gornyi2014,Goldstein2013,Cheianov2013,Altshuler2013,Goldstein2014,Yudson2015,Schmidt2015,Glazman2016,Nagaev2016,Kimme2016,Meir2017,Mani1,Hsu1,Hsu2,Kurilovichi2017,Kurilovichi2018,Yudson2018}.

In contrast to the average current,  shot noise at the helical edge has attracted much less experimental  and theoretical attention so far \cite{Rosenow2013,Khrapai1,Khrapai2,Nagaev2016,VG2017,Nagaev2018}. The shot noise due to backscattering of helical edge electrons via anisotropic exchange (which has to break the conservation of the total $z$-projection of the angular momentum to affect the dc current~\cite{Tanaka2011}) with a local spin $S = \frac{1}{2}$ magnetic moment  has been calculated in Ref.~\cite{VG2017}. The authors of Ref.~\cite{VG2017} studied the so-called \emph{backscattering Fano factor}, $F_{\rm bs}$, which is the ratio between the zero-frequency noise of the backscattering current,  $\mathcal{S}_{\rm bs}$, and the absolute value of the average backscattering current, $|I_{\rm bs}|$, in the limit of large voltage bias $V$. It was found that $F_{\rm bs}$  is bounded between $1$ and $2$, with the extreme values  corresponding to independent backscattering of  single electrons and bunched backscattering of pairs of electrons, respectively. The considerations of Ref. \cite{VG2017} were limited to the case of almost isotropic exchange interaction. This assumption is natural for the model of charge puddles which act as  effective spin-$\frac{1}{2}$ magnetic moments \cite{Goldstein2013,Goldstein2014}. 
However, the spin of a magnetic impurity (MI) can be larger than $\frac{1}{2}$,  e.g., $S = \frac{5}{2}$ for a $\mathrm{Mn}^{2+}$ ion in a HgTe/CdTe quantum well. Moreover, in the case of a ``genuine'' MI the exchange
interaction is strongly anisotropic \cite{Kimme2016,Kurilovichi2017}.

\begin{figure*}[t]
\begin{minipage}{0.875\textwidth}
\begin{minipage}[t]{0.02\textwidth}
\vspace{0cm}
\scriptsize{(a)}
\end{minipage}
\begin{minipage}[t]{0.275\textwidth}
\vspace{0cm}
 \includegraphics[width=\textwidth]{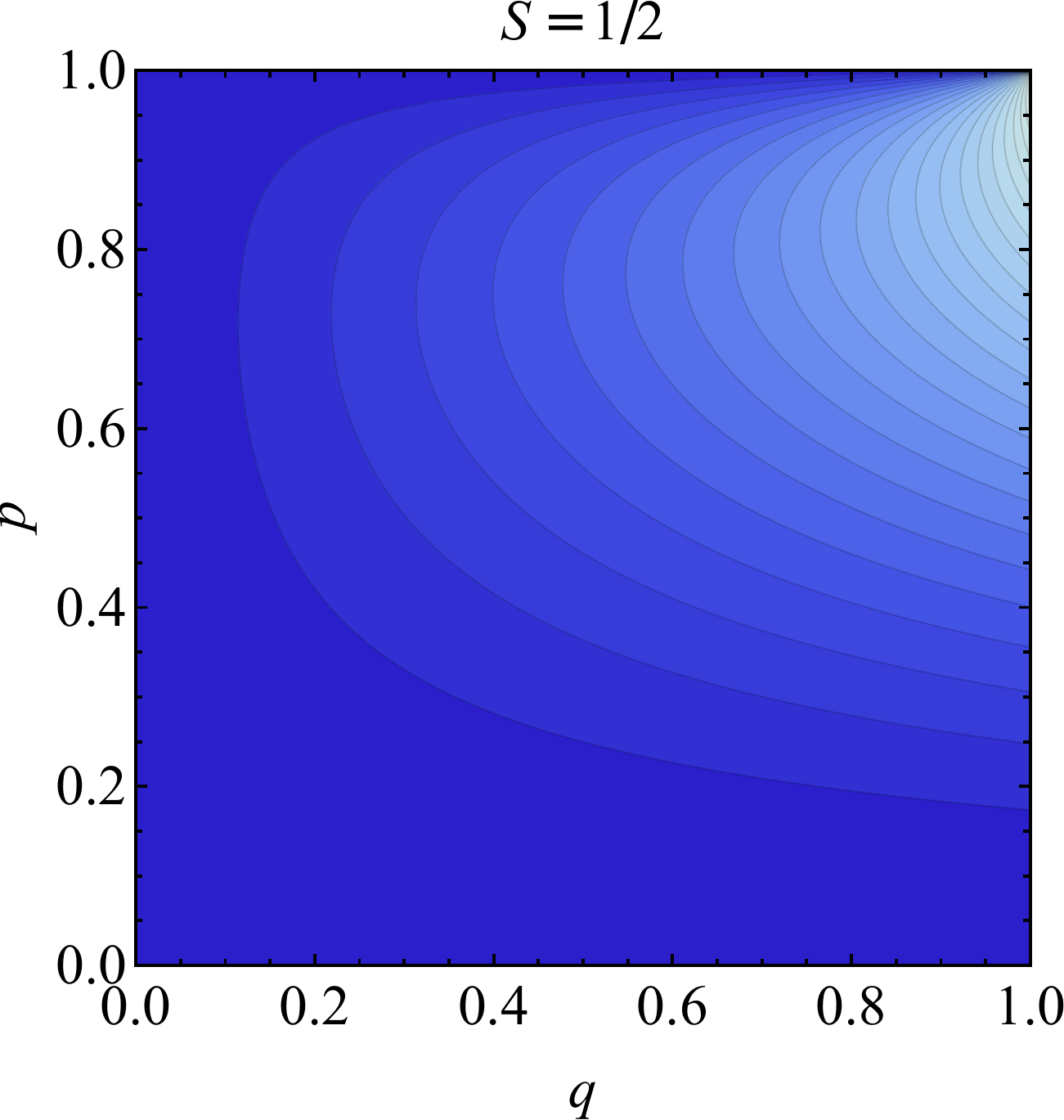}
\end{minipage}
\begin{minipage}[t]{0.02\textwidth}
\vspace{0cm}
\scriptsize{(b)}
\end{minipage}
\begin{minipage}[t]{0.275\textwidth}
\vspace{0cm}
 \includegraphics[width=\textwidth]{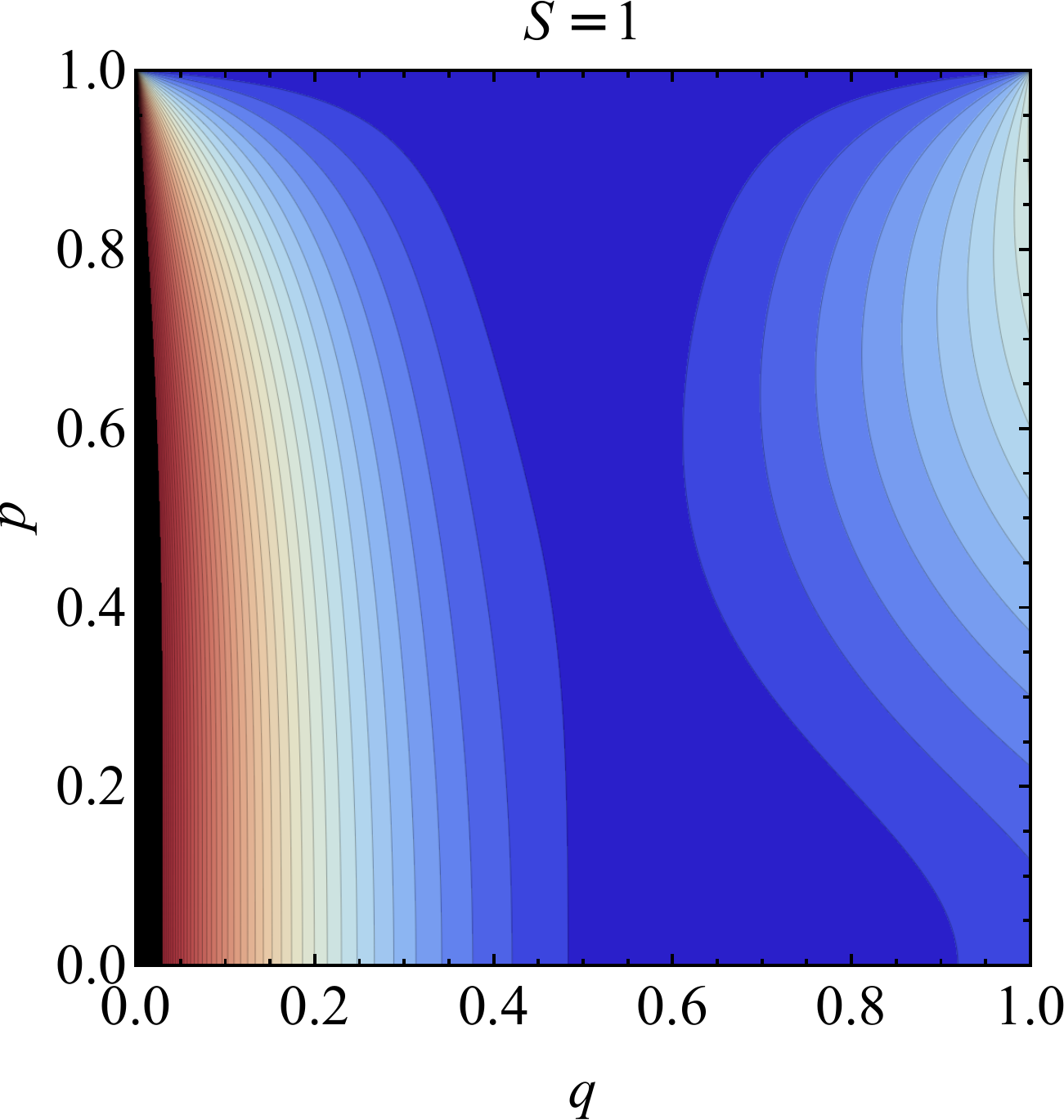}
\end{minipage}
\begin{minipage}[t]{0.02\textwidth}
\vspace{0cm}
\scriptsize{(c)}
\end{minipage}
\begin{minipage}[t]{0.275\textwidth}
\vspace{0cm}
 \includegraphics[width=\textwidth]{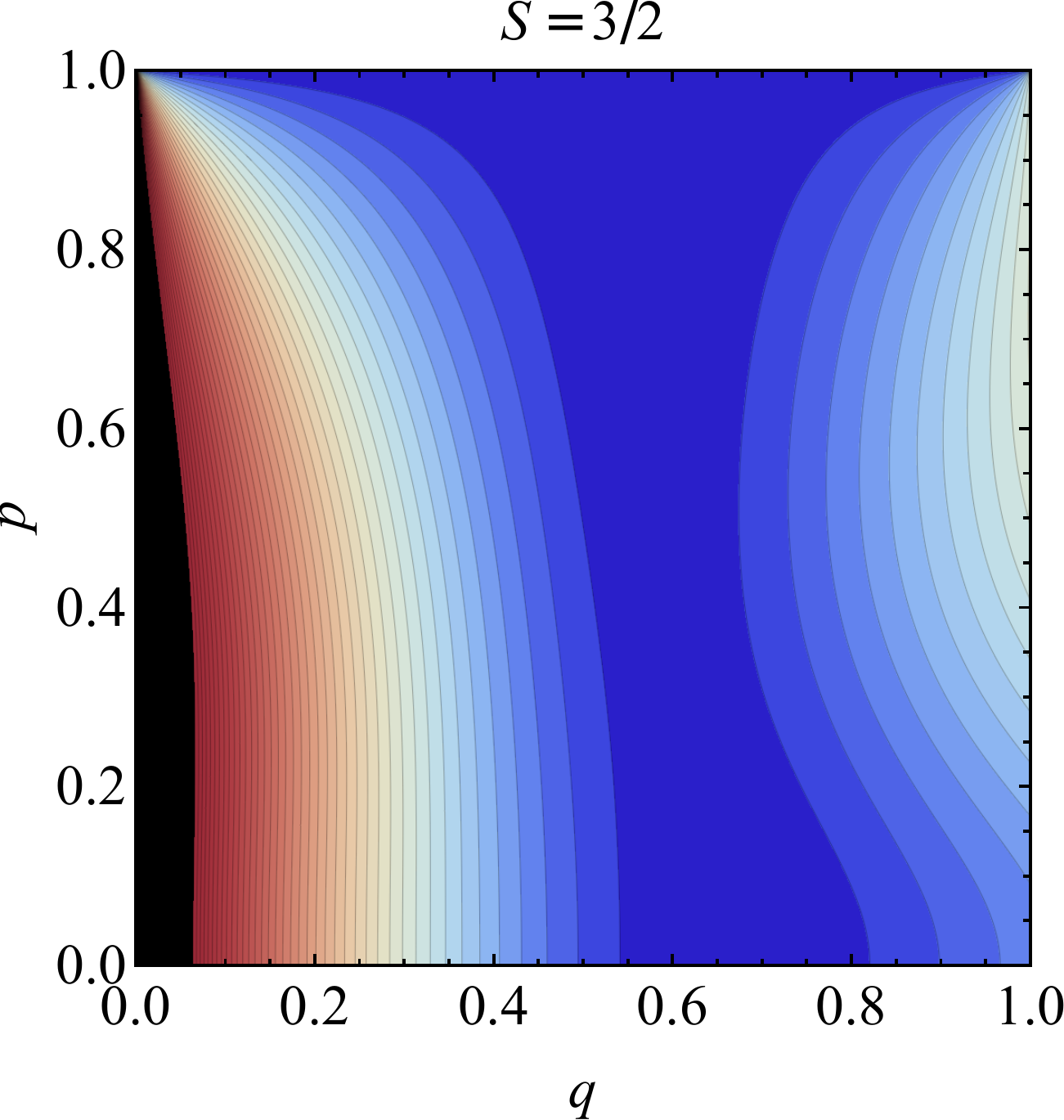}
\end{minipage}%
\vspace{0.5cm}
\linebreak
\begin{minipage}[t]{0.02\textwidth}
\vspace{0cm}
\scriptsize{(d)}
\end{minipage}
\begin{minipage}[t]{0.275\textwidth}
\vspace{0cm}
 \includegraphics[width=\textwidth]{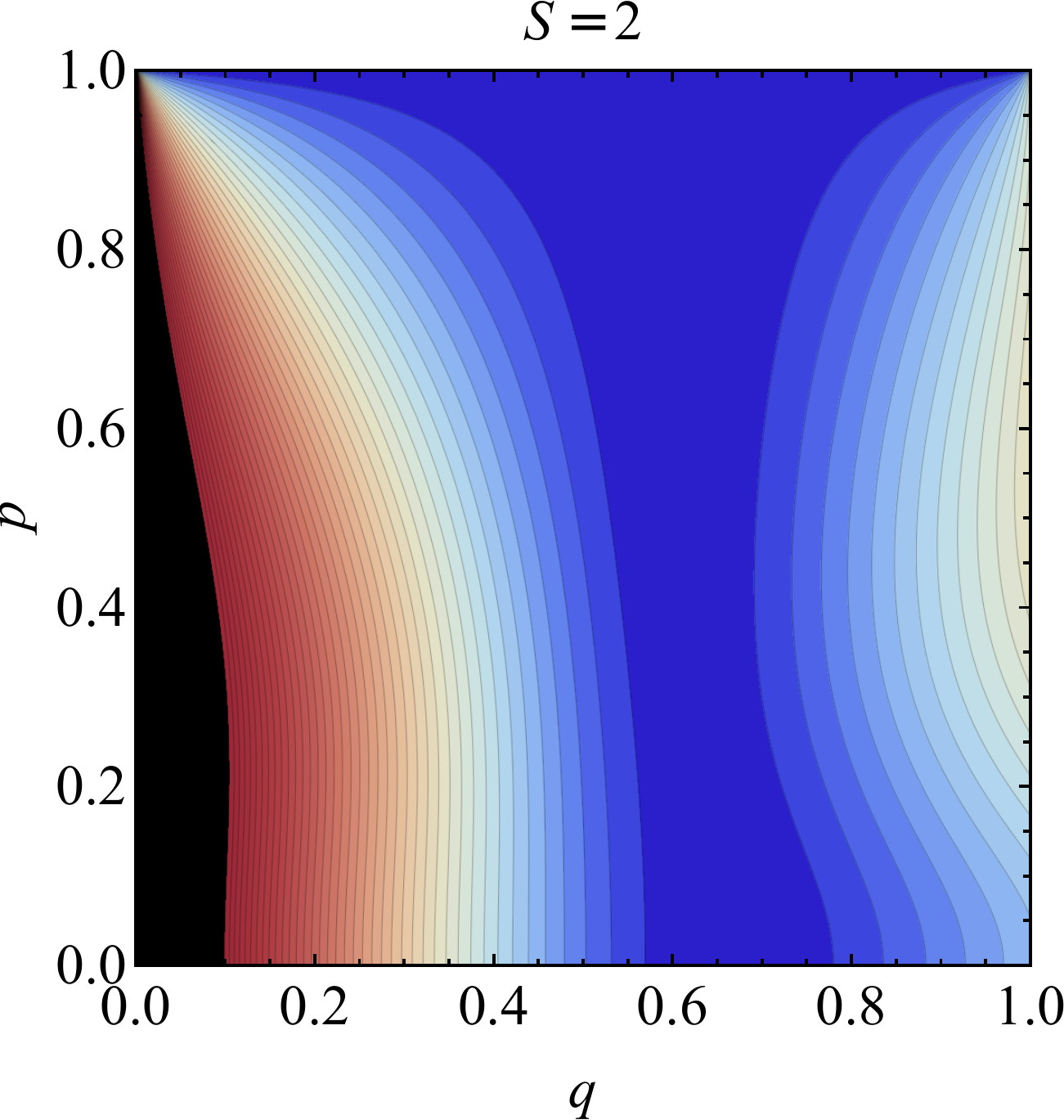}
\end{minipage}
\begin{minipage}[t]{0.02\textwidth}
\vspace{0cm}
\scriptsize{(e)}
\end{minipage}
\begin{minipage}[t]{0.275\textwidth}
\vspace{0cm}
 \includegraphics[width=\textwidth]{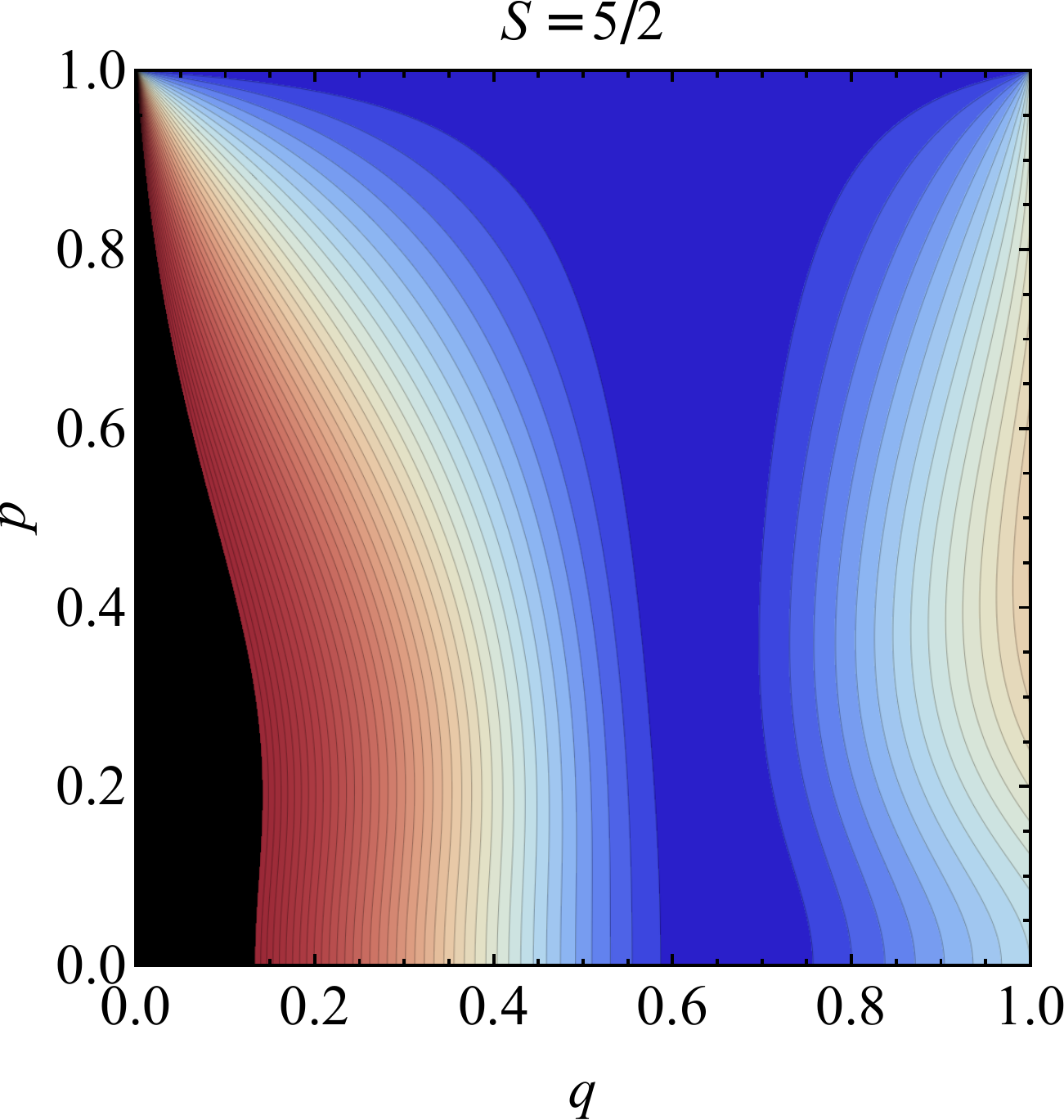}
\end{minipage}
\begin{minipage}[t]{0.02\textwidth}
\vspace{0cm}
\scriptsize{(f)}
\end{minipage}
\begin{minipage}[t]{0.275\textwidth}
\vspace{0cm}
 \includegraphics[width=\textwidth]{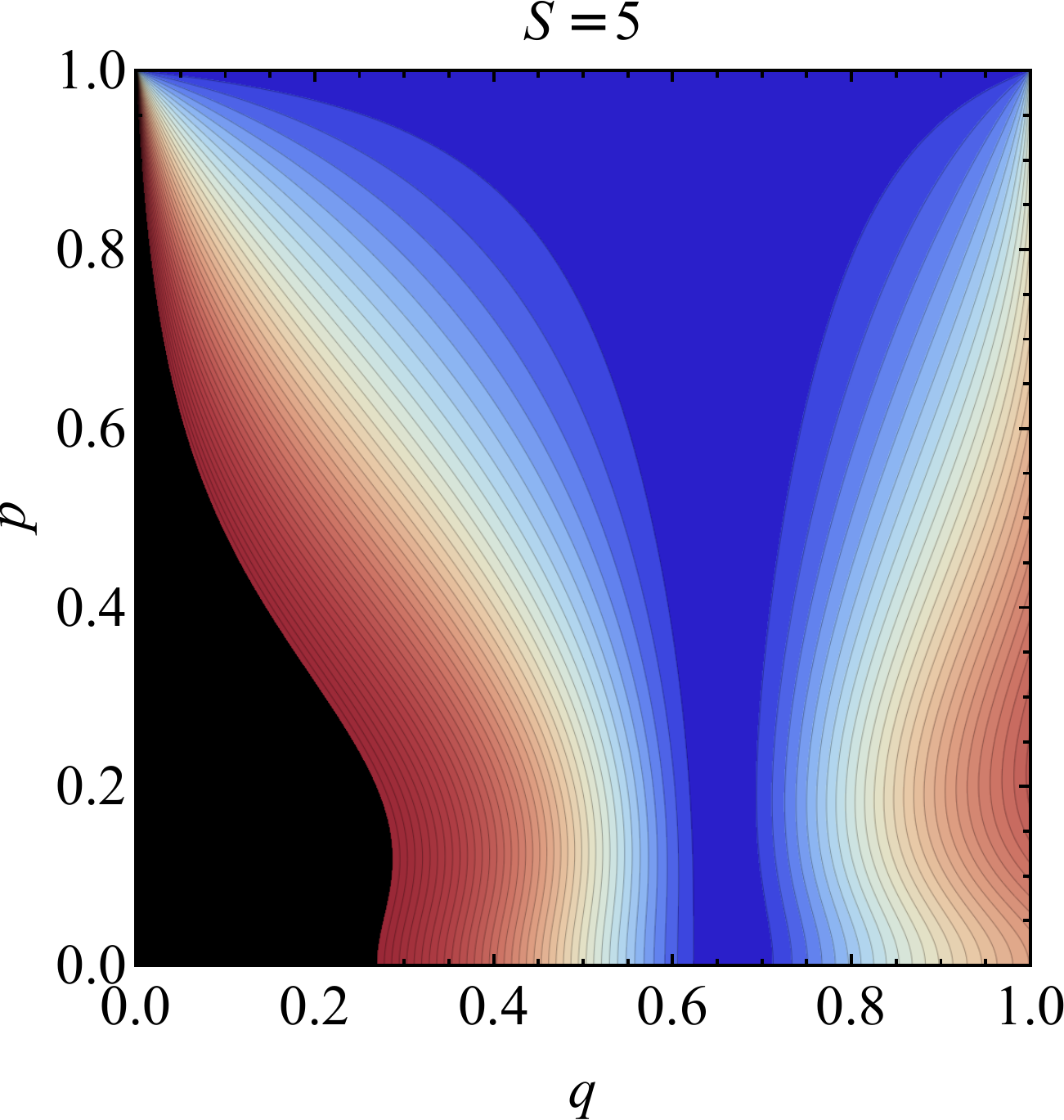}
\end{minipage}
\end{minipage}
\begin{minipage}[c]{0.10\textwidth}
\hspace{-1.5cm}
\includegraphics[width=0.275\textwidth]{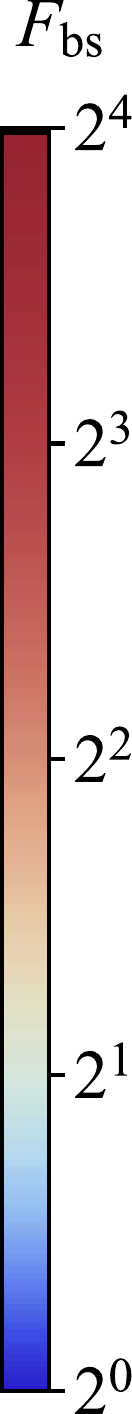}
\end{minipage}
\caption{(Color online) 
The backscattering Fano factor as a function of $q$ and $p$ for different values of the spin: 
(a)~$S=1/2$, (b)~$S=1$, (c)~$S=3/2$, (d)~$S=2$, (e)~$S=5/2$, (f)~$S=5$. Uniform black color corresponds to $F_\mathrm{bs}>2^4$. For $S=1/2$ $F_\mathrm{bs}$ is bounded between 1 and 2. For $S>1/2$ $F_\mathrm{bs}$ diverges in the limit $q\to 0$, except for the line $p=1$, at which $F_\mathrm{bs}=1$. \color{black} 
}
\label{Figure-density-plot}
\end{figure*}

In this Letter we study the backscattering shot noise at the edge of a 2D topological insulator mediated by the presence of a single quantum  MI. We assume that the impurity is of an arbitrary spin $S$ and the exchange interaction matrix is of a general form. We find the
backscattering Fano factor analytically, cf.~Eq.~\eqref{eq:mainresult}. Strikingly, for any $S > \frac{1}{2}$ it is not bounded from above, cf.~Eqs.~\eqref{eq:asymp:p0} and \eqref{eq:asymp:p1}; $F_{\rm bs}>2$  over a wide parameter range, see Fig.~\ref{Figure-density-plot}. 
This implies that a dynamical magnetic moment with $S > \frac{1}{2}$ can bunch helical electrons together. Here, in a significant parameter range, for each value {of the impurity spin projection} $S_z$ electrons are backscattered with a rate $\propto S_z^2$, while $S_z$ {itself} changes slowly{. This results} in a modulation of the backscattering events into long correlated pulses (Fig.~\ref{Figure-refl}a).
For $S = \frac{1}{2}$ this effect is absent (Fig.~\ref{Figure-refl}b), and we find a concise exact expression for $F_\mathrm{bs}$  {proving} rigorously that $1\leqslant F_\mathrm{bs}\leqslant 2$, cf.~Eq.~\eqref{eq:spin12}. Our results elucidate an important facet of the dichotomy between topological properties and electronic correlations in one-dimensional edges \cite{Thomale2019}, accounting for mechanisms that break topological protection against backscattering.

\textsf{Model.}\ --- Helical edge electrons coupled to a MI
are described by the following Hamiltonian (we use units with $\hbar=k_B=-e=1$):
\begin{equation}
\label{eq: tot-H}
H=H_{\mathrm{e}}+H_{\mathrm{e-i}}, \quad H_{\mathrm{e}}= i v\int dy\, \Psi^\dagger(y) \sigma_z \partial_y \Psi(y) ,
\end{equation}
where $\Psi^\dagger$ ($\Psi$) is the creation (annihilation) operator of the edge electrons with velocity $v$. The Pauli matrices $\sigma_{x,y,z}$ act in the spin basis of the edge states. 
The exchange electron-impurity interaction is assumed to be local:
\begin{equation}
\label{eq: e-imp}
H_{\mathrm{e-i}}=\frac{1}{\nu}\mathcal{J}_{ij} S_i s_j(y_{0}),\quad s_j(y) = \frac{1}{2}\Psi^\dagger (y) \sigma_j \Psi (y) .
\end{equation}
Here $\nu = 1/\left( 2\pi v \right)$ is the density of states per one edge mode, 
operator $S_i$ denotes the $i$-th component of the impurity spin, and the couplings $\mathcal{J}_{ij}$ are dimensionless and real.  We stress that in general the exchange interaction \eqref{eq: e-imp} is strongly anisotropic and violates the conservation of the total $z$-projection of the angular momentum of the system \cite{Otten2013,Kimme2016,Kurilovichi2017}. The latter violation is required to generate persistent backscattering of helical electrons. We assume that the coupling constants are small, $|\mathcal{J}_{ij}| \ll 1$, and we neglect the local anisotropy $H_{\mathrm{anis}}=\mathcal{D}_{kp}S_kS_p$ of the MI spin which is justified at $|\mathcal{D}_{kp}| \ll \max\{\mathcal{J}_{ij}^2 T, |\mathcal{J}_{ij}|V\}$ \cite{Kurilovichi2018}. In the absence of the local anisotropy we can rotate the spin basis for $S_i$ bringing the exchange matrix $\mathcal{J}_{ij}$ to a lower triangular form.  We thus assume hereinafter that $\mathcal{J}_{xy}=\mathcal{J}_{xz}=\mathcal{J}_{yz}=0$. In addition we ensure that $\mathcal{J}_{xx}\mathcal{J}_{yy}>0$ with a proper rotation.

\textsf{Cumulant generating function.}\ ---  The average backscattering current and its zero frequency noise can be extracted from the statistics of the number of electrons backscattered off a MI during a large time interval $t$: $\Delta N(t) = \Sigma_z(t) - \Sigma_z$, where $\Sigma_z(t) = e^{i H t} \Sigma_z e^{-i Ht}$ and $\Sigma_z= \int dy s_z(y)$. The cumulant generating function for $\Delta N$ can be written as 
$G(\lambda, t) = \ln \Tr [e^{i\lambda \Sigma_z(t)} e^{-i\lambda \Sigma_z} \rho(0)]$, where $\rho(0)$ stands for the initial density matrix of the full system \cite{Esposito}. It is convenient to write $G(\lambda,t) = \ln \Tr \rho^{(\lambda)}(t)$ where  $\rho^{(\lambda)}(t) = e^{-i H^{(\lambda)} t}\rho(0)  e^{i H^{(-\lambda)} t}$ is the generalized density matrix of the system at time $t$ and $H^{(\lambda)} = e^{i \lambda \Sigma_z/2} H e^{-i \lambda \Sigma_z/2}$. Tracing out the degrees of freedom of the helical  electrons, we obtain $G(\lambda,t) = \ln \Tr_S \rho_S^{(\lambda)}(t)$, where $\rho_S^{(\lambda)}(t)$ denotes the reduced generalized density matrix of the impurity. 

\textsf{Generalized master equation.}\ --- In order to find 
 $G(\lambda,t)$ we derive a generalized Gorini-Kossakowski-Sudarshan-Lindblad equation, which governs the time evolution of $\rho_S^{(\lambda)}(t)$ (see Supplemental Material~\cite{SM}):
\begin{equation}
\frac{d\rho_S^{(\lambda)}}{dt} \!=\!
-i [H_{\rm e-i}^{\rm mf},\rho_S^{(\lambda)}]
\!+\! \eta_{jk}^{(\lambda)} S_j \rho_S^{(\lambda)} S_k -
\frac{\eta_{jk}^{(0)}}{2}\{\rho_S^{(\lambda)},S_k S_j\} .
\label{eq:qms}
\end{equation}
Here $H_{\rm e-i}^{\rm mf} = \mathcal{J}_{zz}\langle s_z\rangle S_z/\nu$ is the mean-field part of $H_{\rm e-i}$ with
the average non-equilibrium spin density $\langle s_z\rangle = \nu V/2$.
{Additionally, we have introduced $\eta_{jk}^{(\lambda)} = \pi T (\mathcal{J} \Pi_V^{(\lambda)} \mathcal{J}^T)_{jk}$, where 
\begin{equation}
 \Pi_V^{(\lambda)} = \begin{pmatrix}
 f_\lambda^+\left(V/T\right)
& -i f_\lambda^-\left(V/T\right) & 0 \\
i f_\lambda^-\left(V/T\right)& f_\lambda^+\left(V/T\right)& 0 \\
0 & 0& 1
\end{pmatrix} 
\end{equation}
and $f_\lambda^\pm(x) = \frac{x}{2}\left(e^{-i\lambda}e^x \pm e^{i\lambda} \right)/\left(e^x-1\right)$.

Below we focus on the regime $V\gg T$. The first term on the r.h.s.~of Eq.~\eqref{eq:qms} is then much larger than the other two terms. Consequently, one may implement the rotating wave approximation to simplify Eq.~\eqref{eq:qms}. Within its framework $\rho_S^{(\lambda)}$ is diagonal in the eigenbasis of  $H_{\rm e-i}^{\rm mf}$, i.e., of $S_z$. Denoting the impurity state with $S_z = m$ as $| m \rangle$ ($m=S,...,-S$) we obtain a classical master equation for the occupation numbers
\begin{equation}
\frac{d}{d t} \langle m |\rho_S^{(\lambda)}|m\rangle =\sum_{m^\prime=-S}^S
\mathcal{L}^{(\lambda)}_{mm^\prime} \langle m^\prime |\rho_S^{(\lambda)}|m^\prime\rangle.
\label{eq:cme}
\end{equation}
Here $\mathcal{L}^{(\lambda)}$ is a $(2S+1)\times(2S+1)$ tridiagonal matrix. The tridiagonal form indicates that $S_z$ changes by not more than unity in each elementary scattering process. Nonzero elements of $\mathcal{L}^{(\lambda)}$ are given by $\mathcal{L}_{m+1,m}^{(\lambda)}=e^{-i\lambda} \eta_+ [S(S+1)-m(m+1)]/4$, 
$\mathcal{L}_{m,m+1}^{(\lambda)} = (\eta_-/\eta_+)\mathcal{L}_{m+1,m}^{(\lambda)}$, and
$\mathcal{L}_{mm}^{(\lambda)} = - e^{i\lambda}\mathcal{L}_{m+1,m}^{(\lambda)}-
e^{i\lambda}\mathcal{L}_{m-1,m}^{(\lambda)} +(e^{-i\lambda}-1) \eta_{zz}^{(0)} m^2$, where $\eta_\pm=\eta_{xx}^{(0)}+\eta_{yy}^{(0)}\pm i (\eta_{xy}^{(0)}-\eta_{yx}^{(0)})$.  It is worthwhile to note that by Eq.~\eqref{eq:cme}, the characteristic function of $\langle m |\rho_S^{(\lambda)}|m\rangle$ obeys the Heun equation~\cite{DMLF}.

\textsf{Results.}\ --- At $\lambda = 0$ Eq.~\eqref{eq:cme} describes the time evolution of populations of the impurity energy levels. Through this equation we establish that the steady state density matrix of the impurity at $V\gg T$ is given by
\begin{equation}
\label{eq: p definition}
\rho^{(0)}_{S,\mathrm{st}} \sim \left(\frac{1+p}{1-p}\right)^{S_z},\quad p = \frac{2{\mathcal{J}_{xx} \mathcal{J}_{yy}}}{\mathcal{J}_{xx}^2+\mathcal{J}_{yy}^2+\mathcal{J}_{yx}^2},
\end{equation}
The dimensionless parameter $p$ determines the polarization of the impurity, i.e., for $p=1$ only the state $S_z = S$ is occupied, whereas for $p=0$ all levels are equally populated. Physically, at $p=1$, $\mathcal{J}_{xx}=\mathcal{J}_{yy}$, $\mathcal{J}_{yx} = 0$, and
the impurity spin can be flipped down only by backscattering an  edge electron carrying spin down. We note, though, that at large voltage the current is carried mainly by spin-up
electrons. Thus, a steady state of the impurity is established in which $S_z = S$ with essentially unit probability. At $p<1$ the impurity spin can be flipped down by electrons with spin up, resulting in depolarization of the impurity.  We stress that $\mathcal{J}_{zx}$ and $\mathcal{J}_{zy}$ do not enter into the expression \eqref{eq: p definition} for $p$ because the corresponding terms in the Hamiltonian do not induce impurity spin flips.

To express $I_{\rm bs}$ and $\mathcal{S}_{\rm bs}$ in a compact form we introduce two parameters:
\begin{equation}
g=\mathcal{J}_{xx}^2+\mathcal{J}_{yy}^2+\mathcal{J}_{yx}^2+\mathcal{J}_{zx}^2+\mathcal{J}_{zy}^2,
\hspace{0.325cm} q =1 - \frac{\mathcal{J}_{zx}^2+\mathcal{J}_{zy}^2}{g} .
\end{equation}
Then we find that 
$\eta_\pm = \pi g V (1\pm p) q/2$ and $\eta^{(0)}_{zz} = \pi g V (1-q)/2$. Notice that $0<p, q \leqslant 1$ and~$g\ll 1$.

The average backscattering current can be found as 
$I_{\rm bs} = \langle \Delta N \rangle /t = -(i/t) \partial G(\lambda,t)/\partial \lambda$, where the limits $t\to \infty$ and $\lambda\to 0$ are assumed. Solving Eq.~\eqref{eq:cme} within the first order perturbation theory in $\lambda$, we find 
\begin{equation}
I_{\rm bs} =- \frac{\pi g V}{4} \langle R(S_z)\rangle ,
\end{equation}
where $R(S_z)= q S(S+1)- q p S_z+
(2-3q) S_z^2$ and $\langle ... \rangle = \mathrm{Tr} \bigl( ...\:\rho^{(0)}_{S,\mathrm{st}}\bigr)$.
We note that $\langle R(S_z)\rangle > 0$, hence $I_{\rm bs}$ is  
negative.

The backscattering current noise at zero frequency is given by the second cumulant of $\Delta N$ as $\mathcal{S}_{\rm bs} = 
\llangle (\Delta N)^2\rrangle /t = - t^{-1} \partial^2 G(\lambda, t)/\partial \lambda^2$ at $t\to \infty$ and $\lambda \to 0$. In order to compute $\mathcal{S}_{\rm bs}$ from Eq.~\eqref{eq:cme} we employ second order perturbation theory in $\lambda$~\cite{SM}.
The noise can be 
written as
$\mathcal{S}_{\rm bs} = F_{\rm bs} |I_{\rm bs}|$, where the backscattering Fano factor
reads
\begin{gather}
F_\mathrm{bs} = 1 + \frac{4}{q(1-p)}\sum_{n=1}^{2S} \frac{\langle P_n [R(S_z) - \langle R(S_z)\rangle]\rangle^2}{n(2S+1-n)\langle R(S_z)\rangle\mu_n}.
\label{eq:mainresult}
\end{gather}
Here $P_n = \sum_{m=S-n+1}^S | m\rangle\langle m|$ is a projector on the subspace of $n$ impurity states with largest $S_z$ projection and $\mu_n = \langle S+1-n|\rho_{S,\mathrm{st}}^{(0)}|S+1-n\rangle$. 
Notice that Eq.~\eqref{eq:mainresult} implies $F_{\rm bs}\geqslant 1$. So far, we considered the model of noninteracting edge states. Accounting for electron-electron interaction results only in the common factor for $I_{\rm bs}$ and $\mathcal{S}_{\rm bs}$ that leaves $F_{\rm bs}$ intact~\cite{SM}.

\begin{figure}[t]
\begin{minipage}[t]{0.015\textwidth}
\vspace{0cm}
\scriptsize{(a)}
\end{minipage}
\begin{minipage}[t]{0.205\textwidth}
\vspace{0cm}
 \includegraphics[width=\textwidth]{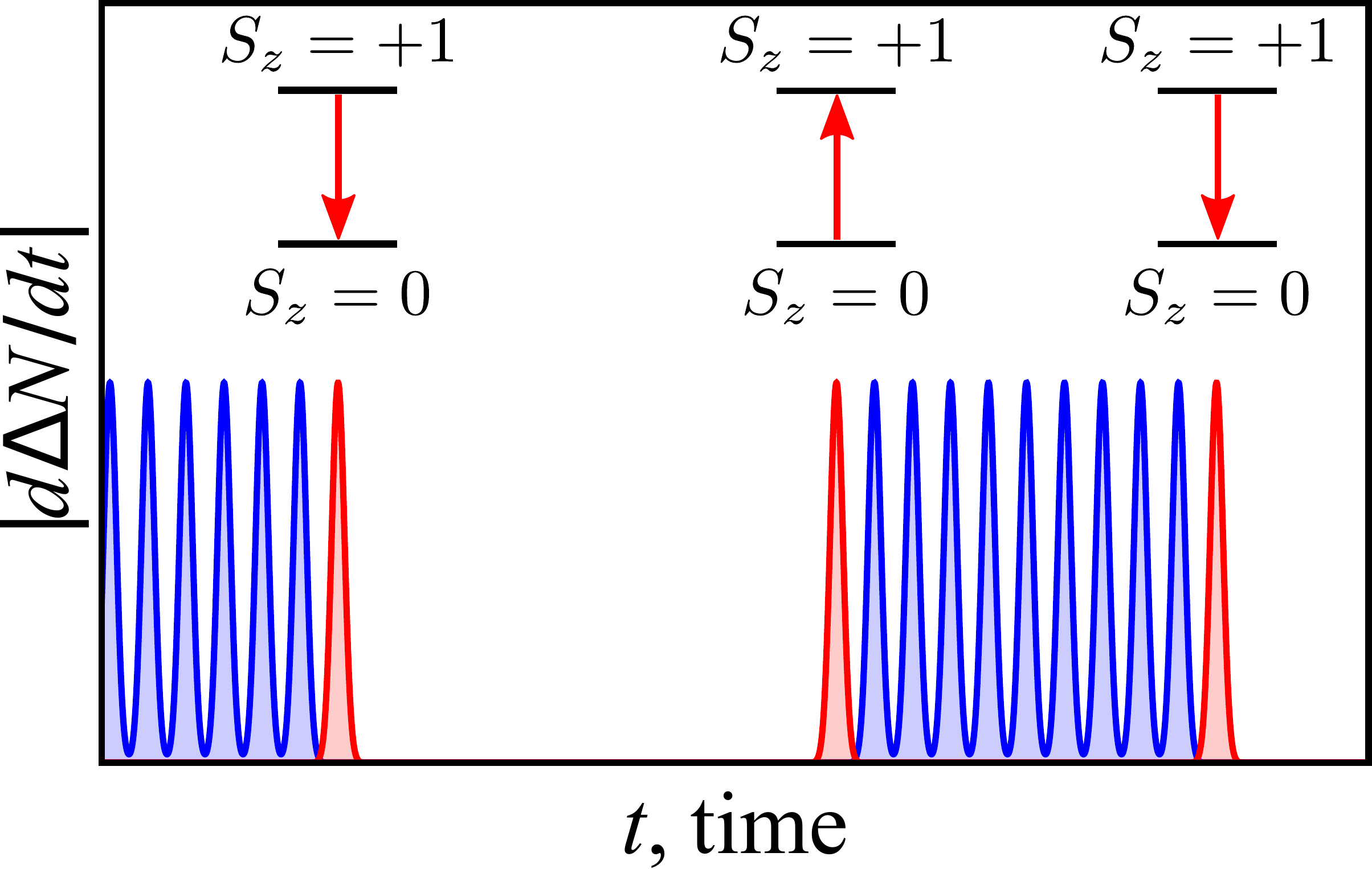}
\end{minipage}
\hspace{0.25cm}
\begin{minipage}[t]{0.015\textwidth}
\vspace{0cm}
\scriptsize{(b)}
\end{minipage}
\begin{minipage}[t]{0.205\textwidth}
\vspace{0cm}
 \includegraphics[width=\textwidth]{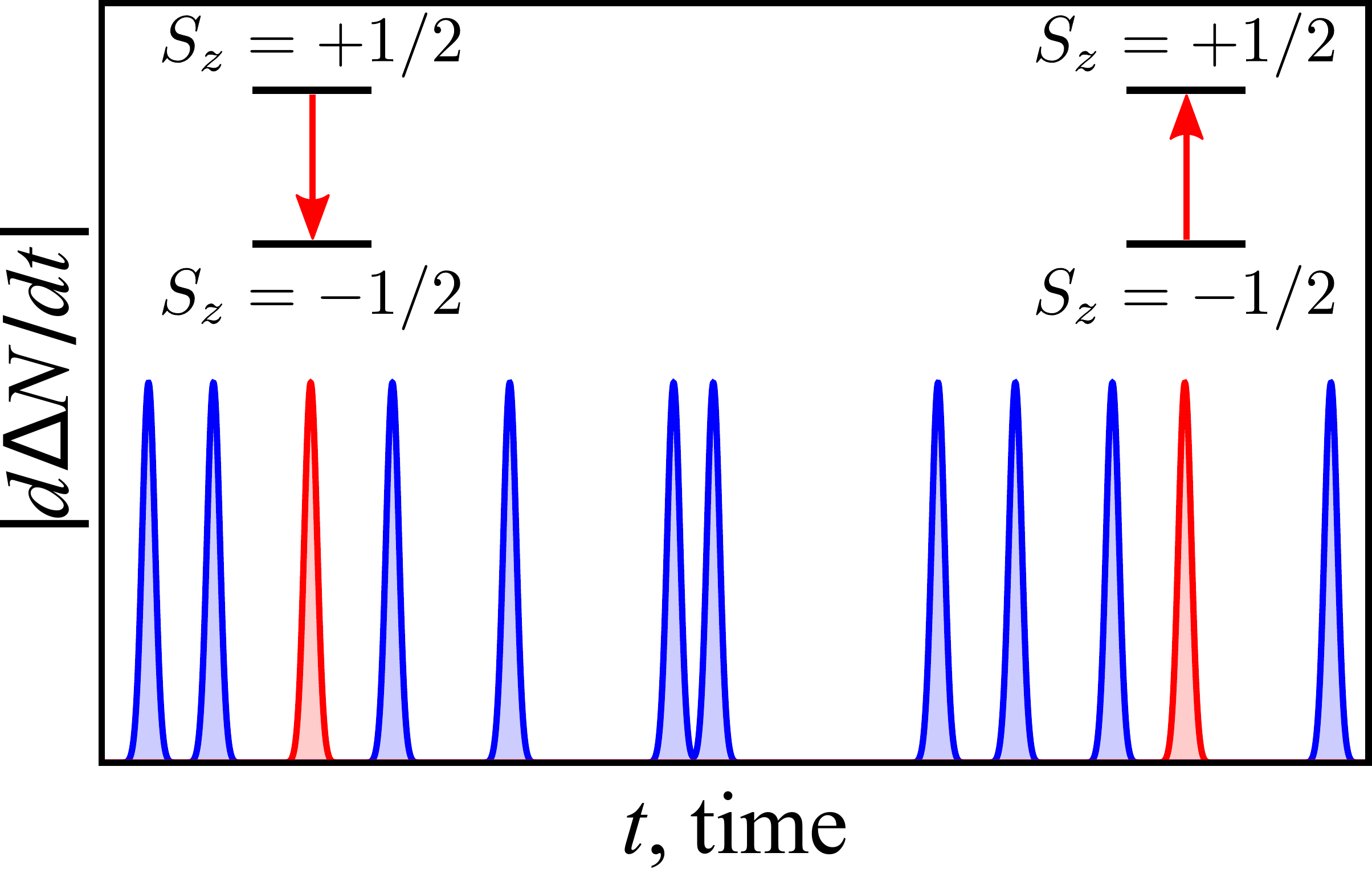}
\end{minipage}%
\vspace{0cm}
\begin{minipage}[t]{0.015\textwidth}
\vspace{0cm}
\scriptsize{(c)}
\end{minipage}
\begin{minipage}[t]{0.205\textwidth}
\vspace{0cm}
 \includegraphics[width=\textwidth]{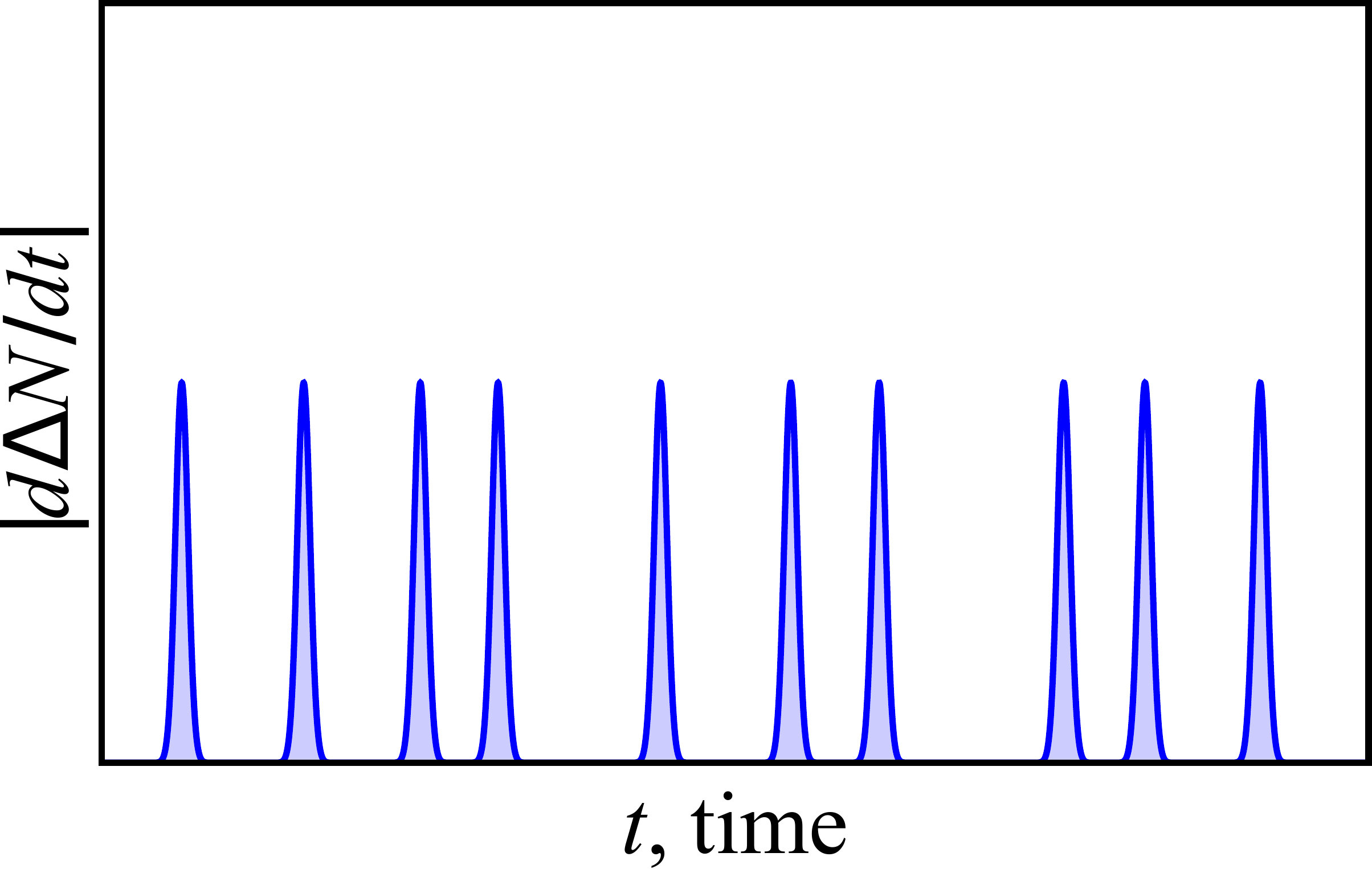}
\end{minipage}
\hspace{0.25cm}
\begin{minipage}[t]{0.015\textwidth}
\vspace{0cm}
\scriptsize{(d)}
\end{minipage}
\begin{minipage}[t]{0.205\textwidth}
\vspace{0cm}
 \includegraphics[width=\textwidth]{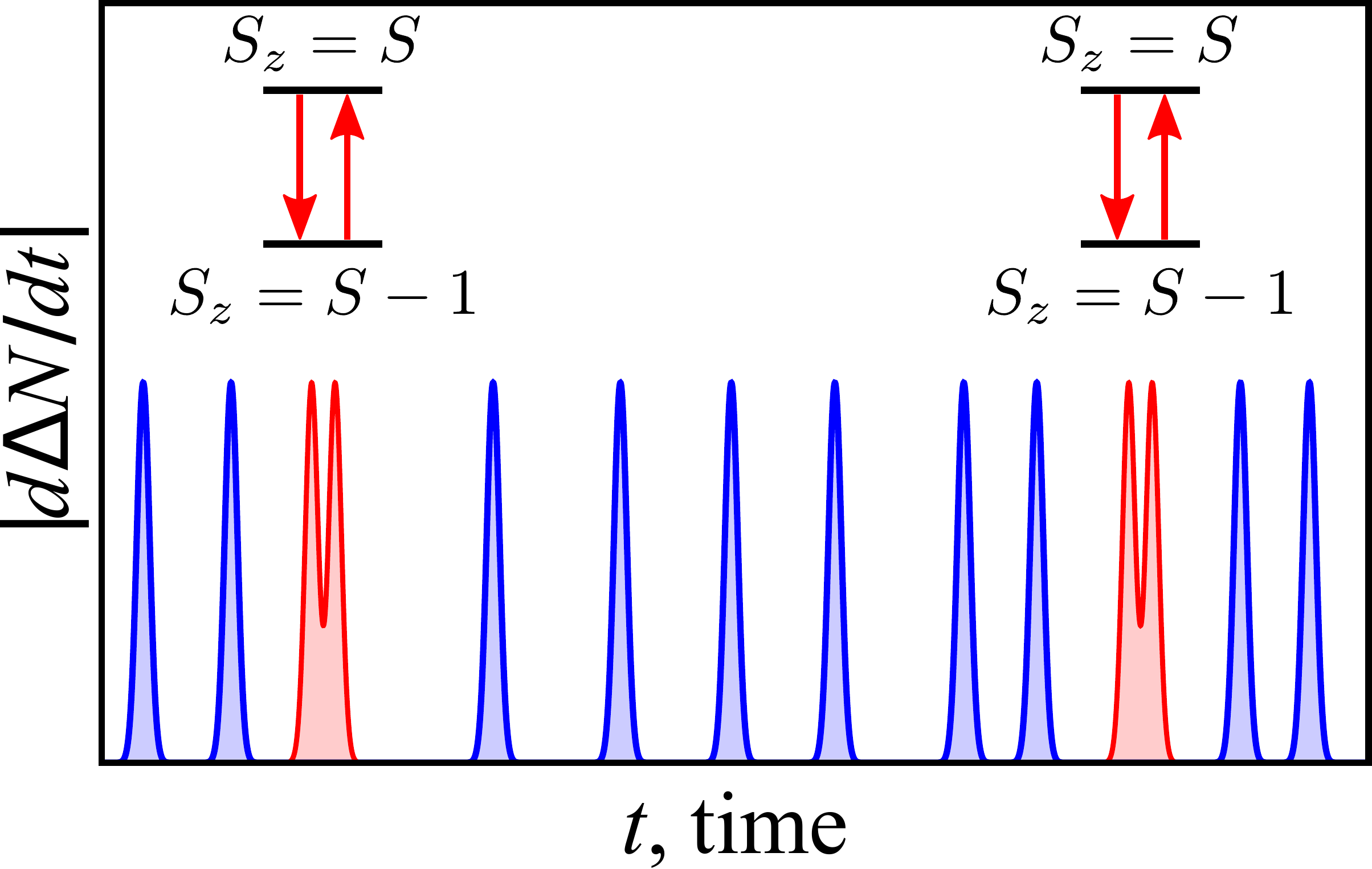}
\end{minipage}%
\caption{(Color online) Sketches of the backscattering current as a function of time in different regimes: (a) $q\ll 1$ and $S=1$; (b) 
$q\ll 1$ and $S=1/2$; (c) $p=1$; (d) $1-p\ll 1$.  Red and blue peaks correspond to backscattering processes with and without the impurity flips, respectively. Transitions between impurity levels are depicted above each spin-flip process.
}
\label{Figure-refl}
\end{figure}

The most striking feature of Eq.~\eqref{eq:mainresult} is the divergence at $q\to 0$ for $0\leqslant p < 1$ (cf.~Eq.~\eqref{eq:asymp:p1}). It indicates that in general the Fano factor is \emph{unrestricted} from above. The only \emph{exception} is the case of $S={1}/{2}$, for which Eq.~\eqref{eq:mainresult} gives
\begin{equation}
F_{\rm bs}(S=1/2) = (1-q p^4)/(1-q p^2).
 \label{eq:spin12}
\end{equation}
This expression indicates that $F_\mathrm{bs}$ is restricted to the range between $1$ and $2$ for $S={1}/{2}$.   Eq.~\eqref{eq:spin12} extends the $p,q\to 1$ result of Ref. \cite{VG2017} to arbitrary values of $p$ and~$q$. 

The divergence of $F_{\rm bs}$ at $q\to 0$  for impurities with $S>1/2$  can be explained on physical grounds. For the sake of simplicity we first consider  $S = 1$. The inequality $q\ll 1$ implies $|\mathcal{J}_{zx}|, |\mathcal{J}_{zy}| \gg |\mathcal{J}_{xx}|, |\mathcal{J}_{yy}|, |\mathcal{J}_{yx}|$ and, therefore, the backscattering  predominantly happens without spin flips of the impurity. By Fermi's golden rule, the rate of such reflection processes is proportional to $S_z^2$, rendering 
the backscattering current, $d \Delta N/dt$, very sensitive to the spin state of the impurity.
The processes associated with $\mathcal{J}_{xx}$, $\mathcal{J}_{yy}$, and $\mathcal{J}_{yx}$ in $H_{\mathrm{e-i}}$ are incapable of producing a significant contribution to $d \Delta N/dt$ on their own}, but they can transfer the impurity from one spin state to another, switching efficient backscattering  on ($S_z = \pm 1$) and off ($S_z = 0$). Consequently, the backscattering current as a function of time looks like a sequence of long pulses, each consisting of a large number (proportional to $1/q$) of backscattered electrons (see Fig.~\ref{Figure-refl}a).  This peculiar bunching of helical  electrons results in $1/q$ divergence of $F_{\rm bs}$.  For $S>1$ the backscattering current looks differently because the reflection intensity remains finite between the pulses. Still, many-electron correlations are present in $d \Delta N/dt$: impurity rarely jumps between states of different $S_z$ changing the intensity of backscattering $\propto S_z^2$.
As $q \rightarrow 0$, impurity backscatters increasingly large number of electrons during its stay in a state with a given $S_z$, which results in the divergent backscattering Fano factor. We note that for $S=1/2$ for both spin states $S_z^2=1/4$.  Because of that $d \Delta N/dt$ has no pulses (see Fig.~\ref{Figure-refl}b) and $F_{\rm bs}$ is not singular at $q\rightarrow 0$.

The exact analytical result \eqref{eq:mainresult}  for $F_\mathrm{bs}$ can be expressed  as a rational function of $p$ and $q$ for any given value of $S$. However, even for $S=1$ such an expression is lengthy. We thus focus on relevant limiting cases below. For an unpolarized impurity, $p=0$, 
we find 
\begin{equation}
F_{\rm bs}(p=0)  = 1+ (2S-1)(2S+3)(2-3q)^2/(45 q) .
\label{eq:asymp:p0}
\end{equation}
In this regime the Fano factor scales as $S^2$ at large $S$.  This fact has a simple physical interpretation. Since for $p=0$ each of the $2S+1$ spin states of the impurity is occupied with the same probability, the dynamics of impurity flips between states with different $S_z$ is diffusive. If the MI starts its motion in a state 
$S_z$, on average $\sim S^2$ transitions occur before $S_z$ returns to its initial value. Therefore, approximately $S^2$ subsequent spin flips of the impurity are correlated. These correlations in the dynamics of the impurity spin are mirrored by the correlations in the electron backscattering and result in the $S^2$ scaling of $F_{\rm bs}$ at small $p$. 

For an almost fully polarized MI, $1-p\ll 1-q$, Eq.~\eqref{eq:mainresult}  yields
\begin{equation}
F_{\rm bs}(p\to1)  =1 + \frac{(1-p)\bigl((2-3q)S+q-1\bigr )^2}{ q(1-q) S^3} .
\label{eq:asymp:p1}
\end{equation}
As follows from Eq.~\eqref{eq:asymp:p1} the Fano factor at $p=1$ is equal to unity for $q<1$. This result is expected since for $p=1$ the spin of a MI is locked to the state $S_z =S$. Therefore, the only allowed backscattering processes occur due to the $\mathcal{J}_{zx}$ and $\mathcal{J}_{zy}$ terms in $H_{\mathrm{e-i}}$, which do not require spin-flips of the impurity to scatter helical electrons. Consequently, the impurity does not keep memory about backscattered electrons, which results in a Poissonian single-electron reflection process with $F_{\rm bs} = 1$ (see Fig.~\ref{Figure-refl}c). 
For $1-p\ll 1$, 
rare two-particle reflections are involved in addition to the single-particle backscattering. They are accompanied by short-time excursions of the impurity spin from the state $S_z = S$ to the state $S_z = S-1$
and lead to the enhancement of the Fano
factor above unity (cf.~Eq.~\eqref{eq:asymp:p1}). In total,
for $1-p\ll 1$ the backscattering of electrons represents 
superposition of the independent single- and two-
particle Poisson processes (see Fig.~\ref{Figure-refl}d).
For large $S$ the deviation of $F_{\rm bs}$ from unity is additionally suppressed by a factor $1/S$ in the considered limit 
[cf.~Eq.~\eqref{eq:limit}].

The behavior of $F_{\rm bs}$ at the point $q=p=1$ is \emph{non-analytical} owing to $I_{\rm bs}=0$. The value of the Fano factor depends on the direction in the $(q,p)$ plane at which this point is approached. 
For a fixed ratio $\frac{1-p}{1-q}$, Eq.~\eqref{eq:mainresult} yields:
\begin{equation}
F_{\rm bs}(q,p\to 1) = \frac{2(1-p)+(1-q)S}{1-p+(1-q)S} .
\end{equation}   

The overall behavior of $F_{\rm bs}(q,p)$ for different values of $S$ is shown in Fig.~\ref{Figure-density-plot}.  For  $S=1/2$ (Fig.~\ref{Figure-density-plot}a) the backscattering Fano factor is bounded by $1\leq F_\mathrm{bs} \leq 2$.  For $S>1/2$ (Figs.~\ref{Figure-density-plot}b - \ref{Figure-density-plot}f) there is a divergence in $F_\mathrm{bs}$ in the vicinity of $q=0$.  
The divergence appears to be
more pronounced as $S$ increases. 
However, this trend breaks  down for large $S$. Eq.~\eqref{eq:mainresult} implies that for $S \gg 1/[p(1-q)(2-3q)]$ the Fano factor behaves as
\begin{equation}
F_\mathrm{bs} = 1 + \frac{1}{2S} \frac{(1-p^2)(2-3q)^2}{qp^3(1-q)}+O(1/S^2),
\label{eq:limit}
\end{equation}
i.e., it gradually decreases and approaches unity as $S$ gets higher. Thus, the limit of large spin corresponds to the backscattering of helical electrons by a \textit{classical} MI.

Along with the line $p=1$ the Fano factor equals unity at the isolated point $(q,p)=(2/3,0)$. This is universal for all $S$ (see Eq.~\eqref{eq:asymp:p0}). At this point the backscattering rate is independent of the impurity state and, therefore,  the backscattering current statistics has a Poissonian single-particle character. The interplay between an $~S^2$ scaling of $F_\mathrm{bs}$ at $p\rightarrow 0$ and the presence of a degenerate point $(q,p)=(2/3,0)$ results in a bottleneck feature in the vicinity of the latter in Figs.~\ref{Figure-density-plot}b - \ref{Figure-density-plot}f. The $1/S$ term in Eq. \eqref{eq:limit} cancels at the line $q=2/3$  and $F_\mathrm{bs}(q=2/3)-1\sim 1/S^3$ for~$p\neq 0$. 

\textsf{Conclusions.}\ ---  To summarize, we have investigated the zero-frequency statistics of the backscattering current induced by a magnetic impurity of an arbitrary spin $S$ located near the edge of a two-dimensional topological insulator. We addressed the limit of large voltage  $|V|\gg \max\{T,|\mathcal{D}_{kp}/ \mathcal{J}_{ij}|\}$ where it is possible to neglect the thermal contribution to the noise, as well as the effect of the local anisotropy of the magnetic impurity. Our analytical solution for the average backscattering current and its zero-frequency noise underscores several striking features:  (i) the dependence of the average backscattering current and noise on the elements of the exchange matrix is determined by three parameters ($g, p, q$) only  instead of a-priory six different parameters (the number of non-zero elements of the exchange matrix $\mathcal{J}_{ij}$); (ii) for $S>1/2$ the backscattering Fano factor can be arbitrary large, diverging in the limit $q\to 0$. This implies that the backscattered electrons can be bunched together in long pulses. Observation of electron bunching is a novel challenge for experimentalists which might shed light on the nature of backscattering at the helical edge, and on how emerging strong correlations can undermine topological protection against backscattering; (iii) for $S=1/2$ the backscattering Fano factor is limited to the range between $1$ and $2$; (iv) the backscattering Fano factor is independent of the electron-electron interaction.

\textsf{Acknowledgements.}\ --- We thank V. Kachorovskii for discussions. Hospitality by Tel Aviv University, the Weizmann Institute of Science, the Landau Institute for Theoretical Physics, and the Karlsruhe Institute of Technology is gratefully acknowledged. The work was partially supported by the programs of the Russian Ministry of Science and Higher Education, 
DFG Research Grant SH 81/3-1, the Italia-Israel QUANTRA, the Israel Ministry of Science and Technology (Contract No.~3-12419),  the Israel Science Foundation (Grant No.~227/15), the German Israeli Foundation (Grant No.~I-1259-303.10), and the US-Israel Binational Science Foundation (Grant No.~2016224), \color{black} and a travel grant by the BASIS Foundation (Russia).




\section*{Supplementary material}

# Supplemental Material for "Unrestricted electron bunching at the helical edge"

Pavel D. Kurilovich,[1] Vladislav D. Kurilovich,[1] Igor S. Burmistrov,[2,3] Yuval Gefen,[4] and Moshe Goldstein[5]

[1]*Department of Physics, Yale University, New Haven, CT 06520, USA*
[2]*L. D. Landau Institute for Theoretical Physics, acad. Semenova av. 1-a, 142432 Chernogolovka, Russia*
[3]*Condensed-matter Physics Laboratory, National Research University Higher School of Economics, 101000 Moscow, Russia*
[4]*Department of Condensed Matter Physics, The Weizmann Institute of Science, Rehovot 76100, Israel*
[5]*Raymond and Beverly Sackler School of Physics and Astronomy, Tel Aviv University, Tel Aviv 6997801, Israel*

In these notes, we (i) present a derivation of the generalized master equation governing the time evolution of the cumulant generating function of the number of backscattered electrons (Eq. (3) of the main text) and (ii) obtain analytical expressions for the backscattering current and the backscattering Fano factor (Eqs. (8) and (9) of the main text, respectively). We also demonstrate that the backscattering Fano factor is insensitive to the presence of electron-electron interaction at the helical edge.

## S.I. DERIVATION OF THE GENERALIZED MASTER EQUATION

In this section we derive a generalized Gorini-Kossakowski-Sudarshan-Lindblad equation that describes the time evolution of the reduced generalized density matrix [S1] of a magnetic impurity spin, $\rho_S^{(\lambda)}(t)$, i.e., Eq. (3) of the main text. Knowledge of $\rho_S^{(\lambda)}(t)$ is required to calculate the cumulant generating function $G(\lambda, t)$ of the number of electrons $\Delta N(t)$ backscattered by the magnetic impurity in a time interval $t$ (we use units where $\hbar = k_B = -e = 1$):

$$G(\lambda, t) = \ln \mathrm{Tr}_S \, \rho_S^{(\lambda)}(t), \quad \rho_S^{(\lambda)}(t) = \mathrm{Tr}_e \, e^{-iH^{(\lambda)}t} \rho(0) e^{iH^{(-\lambda)}t}. \tag{S1}$$

Here $\mathrm{Tr}_S$ and $\mathrm{Tr}_e$ are the traces over the impurity spin states and over the edge electronic degrees of freedom, respectively, $H = H_e + H_{e-i}$ is the full Hamiltonian of the system (the definitions of $H_e$ and $H_{e-i}$ are provided in Eqs. (1) and (2) of the main text),

$$H^{(\lambda)} = e^{i\Sigma_z \lambda/2} H e^{-i\Sigma_z \lambda/2}, \quad \Sigma_z = \int dy \, s_z(y) = \frac{1}{2} \int dy \, \Psi^\dagger(y) \sigma_z \Psi(y), \tag{S2}$$

and $\rho(0) \sim 1_S \otimes \exp\left(-(H_e - \Sigma_z V)/T\right)$ is the initial density matrix of the whole system. In the latter expression $1_S$ is a unity matrix in the impurity spin subspace, $T$ is the temperature, and $V$ is the voltage applied to the helical edge.

To determine the time evolution of $\rho_S^{(\lambda)}(t)$ we first study the dynamics of the generalized density matrix of the whole system:

$$\rho^{(\lambda)}(t) = \exp\left(-iH^{(\lambda)}t\right)\rho(0)\exp\left(iH^{(-\lambda)}t\right). \tag{S3}$$

A Liouville-type equation governing its evolution is as follows

$$\frac{d\rho^{(\lambda)}(t)}{dt} = -i\left[H_e, \rho^{(\lambda)}(t)\right] - ie^{i\Sigma_z \lambda/2} H_{e-i} e^{-i\Sigma_z \lambda/2} \rho^{(\lambda)}(t) + i\rho^{(\lambda)}(t) e^{-i\Sigma_z \lambda/2} H_{e-i} e^{i\Sigma_z \lambda/2}. \tag{S4}$$

Importantly, at finite voltage the electron impurity interaction $H_{e-i} = \mathcal{J}_{ij} S_i s_j(y_0)/\nu$ acquires a mean-field expectation value of $H_{e-i}^{\mathrm{mf}} = \mathcal{J}_{iz} S_i \mathrm{Tr}\left(\rho(0) s_j(y_0)\right)/\nu = \mathcal{J}_{iz} S_i V/2$. We thus divide $H_{e-i} = H_{e-i}^{\mathrm{mf}} + :H_{e-i}:$, where the normal ordering of any operator $\mathcal{O}$ is defined by $:\mathcal{O}:= \mathcal{O} - \mathrm{Tr}\left(\mathcal{O}\rho(0)\right)$. Then by noticing that $[s_j(y_0), \Sigma_z] = i\epsilon_{jzk} s_k(y_0)$ we rewrite Eq. (S4) in the form

$$\frac{d\rho^{(\lambda)}(t)}{dt} = -i\left[H_e + H_{e-i}^{\mathrm{mf}}, \rho^{(\lambda)}(t)\right] - i\frac{1}{\nu} \mathcal{J}_{ij}^{(\lambda)} S_i :s_j(y_0): \rho^{(\lambda)}(t) + i\rho^{(\lambda)}(t) \frac{1}{\nu} \mathcal{J}_{ij}^{(-\lambda)} S_i :s_j(y_0):, \tag{S5}$$

where we introduced

$$\mathcal{J}_{ij}^{(\lambda)} = \mathcal{J}_{ik} R_{kj}^{(\lambda)}, \quad R^{(\lambda)} = \begin{pmatrix} \cos\frac{\lambda}{2} & -\sin\frac{\lambda}{2} & 0 \\ \sin\frac{\lambda}{2} & \cos\frac{\lambda}{2} & 0 \\ 0 & 0 & 1 \end{pmatrix}. \tag{S6}$$

At $\lambda = 0$ Eq. (S5) has a form of a regular Liouville equation and describes the time evolution of the usual density matrix of the system $\rho(t) \equiv \rho^{(\lambda=0)}(t)$. The only modification arising for $\lambda \neq 0$ is the rotation of the exchange matrix in accordance with (S6). Notice that this rotation is in different directions for the second and the third terms on the r.h.s. of Eq. (S5). This implies that the trace of $\rho^{(\lambda)}(t)$ is not conserved at finite $\lambda$ and thus the cumulant generating function $G(\lambda, t)$ is non-trivial.

In order to obtain the equation for the dynamics of the reduced generalized density matrix of the magnetic impurity $\rho_S^{(\lambda)}(t)$, it is necessary to trace out the electronic degrees of freedom in Eq. (S5). To do that, the standard perturbative technique (in $|\mathcal{J}_{ij}| \ll 1$) of the theory of open quantum systems may be employed (for a review, see [S2]). Such derivation for $\lambda = 0$ was conducted recently by the present authors [S3]. Taking into account the modification of the exchange matrix by a finite $\lambda$ and using the results of [S3], we find

$$\frac{d\rho_S^{(\lambda)}}{dt} = -i[H_{\text{e-i}}^{\text{mf}}, \rho_S^{(\lambda)}] + \eta_{jk}^{(\lambda)} S_j \rho_S^{(\lambda)} S_k - \frac{1}{2} \eta_{jk}^{(0)} \{\rho_S^{(\lambda)}, S_k S_j\}. \tag{S7}$$

In this equation $\eta_{jk}^{(\lambda)} = \pi T \big(\mathcal{J}^{(\lambda)} \Pi_V^{(0)} (\mathcal{J}^{(-\lambda)})^T\big)_{jk}$, where $T$ superscript stands for matrix transposition and $\Pi_V^{(0)}$ denotes the Hermitian part of a zero-frequency spin-spin correlation function of the helical edge electrons:

$$\big(\Pi_V^{(0)}\big)_{pq} = \frac{1}{\nu^2} \int_{-\infty}^{+\infty} d\tau \, \text{Tr}\big(:s_q(y_0, \tau): :s_p(y_0, 0): \rho(0)\big), \quad s_q(y_0, \tau) = e^{iH_e \tau} s_q(y_0) e^{-iH_e \tau}. \tag{S8}$$

Performing a straightforward calculation we find

$$\Pi_V^{(0)} = \begin{pmatrix} \frac{V}{2T} \coth \frac{V}{2T} & -i\frac{V}{2T} & 0 \\ i\frac{V}{2T} & \frac{V}{2T} \coth \frac{V}{2T} & 0 \\ 0 & 0 & 1 \end{pmatrix}. \tag{S9}$$

By introducing $\Pi_V^{(\lambda)} = R^{(\lambda)} \Pi_V^{(0)} R^{(\lambda)}$ we can alternatively rewrite $\eta_{jk}^{(\lambda)} = \pi T \big(\mathcal{J} \Pi_V^{(\lambda)} \mathcal{J}^T\big)_{jk}$ and thus obtain Eq. (4) of the main text. Notice that in the shot noise limit of $V \gg T$ the expression for $\Pi_V^{(0)}$ reduces to

$$\Pi_V^{(0)} \approx \frac{V}{2T} \begin{pmatrix} 1 & -i & 0 \\ i & 1 & 0 \\ 0 & 0 & 0 \end{pmatrix}. \tag{S10}$$

For future use we note that the right hand side of Eq. (S7) can be thought of as a linear evolution super-operator acting on the reduced density matrix of the magnetic impurity.

Importantly, while Eq. (S9) is specific for the non-interacting Hamiltonian $H_e$ given by Eq. (1) of the main text, Eqs. (S7) and (S8) are quite general and, therefore, can be readily extended to more complicated models. In particular, electron-electron interactions at the helical edge can be easily incorporated into the above considerations by using the Luttinger liquid formalism [S4]. Their sole effect is to modify the expression for the spin-spin correlation function. If the Luttinger parameter equals $K$ (we assume $K > 1/2$, i.e., that the interaction is not too strong) one finds [S5, S6]

$$\Pi_V^{(0)} = \mathcal{F}_{V,T} \begin{pmatrix} \frac{V}{2T} \coth \frac{V}{2T} & -i\frac{V}{2T} & 0 \\ i\frac{V}{2T} & \frac{V}{2T} \coth \frac{V}{2T} & 0 \\ 0 & 0 & 0 \end{pmatrix} + \frac{1}{K} \begin{pmatrix} 0 & 0 & 0 \\ 0 & 0 & 0 \\ 0 & 0 & 1 \end{pmatrix}, \tag{S11}$$

where

$$\mathcal{F}_{V,T} = \left(\frac{2\pi T a}{u}\right)^{2K-2} B\left(K - i\frac{V}{2\pi T}, K + i\frac{V}{2\pi T}\right) \frac{\sinh \frac{V}{2T}}{\frac{V}{2T}}. \tag{S12}$$

Here $B(x, y)$ is the Euler beta function [S7], $u$ is the excitation velocity (which, in general, differs from $v$ in the non-interacting Hamiltonian), and $a$ is the ultraviolet cutoff length scale. At $K = 1$ the expression (S9) is trivially reproduced. We note that at large voltages, $V \gg T$, $\mathcal{F}_{V,T} \approx (aV/u)^{2K-2}/\Gamma(2K)$ and, therefore, for $K > 1/2$

$$\Pi_V^{(0)} \approx \frac{1}{\Gamma(2K)} \left(\frac{aV}{u}\right)^{2K-2} \frac{V}{2T} \begin{pmatrix} 1 & -i & 0 \\ i & 1 & 0 \\ 0 & 0 & 0 \end{pmatrix}, \tag{S13}$$

where $\Gamma(x)$ is the Gamma function [S7]. Remarkably, the matrix structure of asymptotic expressions (S10) and (S13) is similar. We will show in Sec. S.II that this is the reason for the independence of $F_{\text{bs}}$ on the electron-electron interaction, as stated in the main text.

## S.II. ANALYTICAL EXPRESSIONS FOR THE AVERAGE BACKSCATTERING CURRENT AND FOR THE BACKSCATTERING FANO FACTOR

In the present section we use Eq. (S7) to derive analytical expressions for the average backscattering current and for the backscattering Fano factor (Eqs. (8) and (9) of the main text, respectively). To this end we express the average backscattering current $I_{\rm bs}$ and zero-frequency noise $\mathcal{S}_{\rm bs}$ in terms of the cumulant generating function $G(\lambda, t)$:

$$I_{\rm bs} = \frac{\langle \Delta N \rangle}{t} = -\frac{i}{t} \frac{\partial}{\partial \lambda} G(\lambda, t), \quad \mathcal{S}_{\rm bs} = \frac{\langle\!\langle\!\langle (\Delta N)^2 \rangle\!\rangle\!\rangle}{t} = -\frac{1}{t} \frac{\partial^2}{\partial \lambda^2} G(\lambda, t), \quad t \to \infty, \ \lambda \to 0. \tag{S14}$$

We recall that $G(\lambda, t) = \ln {\rm Tr}_S \rho_S^{(\lambda)}(t)$. Then $G(\lambda, t \to \infty) = {\rm const} + t\epsilon_0(\lambda)$, where $\epsilon_0(\lambda)$ is the eigenvalue of the evolution super-operator featured in the master equation (S7) with the largest real part. At $\lambda = 0$ this eigenvalue vanishes, $\epsilon_0(\lambda = 0) = 0$. Indeed, Eq. (S7) evaluated at $\lambda = 0$ describes the usual dissipative dynamics of the magnetic impurity spin, which has a unique steady state [S3, S8]. At finite but small $\lambda$ the eigenvalue $\epsilon_0(\lambda)$ is pushed away from zero,

$$\epsilon_0(\lambda) = \epsilon_0^{(1)} \frac{\lambda}{1!} + \epsilon_0^{(2)} \frac{\lambda^2}{2!} + O(\lambda^3), \tag{S15}$$

implying that $I_{\rm bs} = -i\epsilon_0^{(1)}$ and $\mathcal{S}_{\rm bs} = -\epsilon_0^{(2)}$. In general, finding the coefficients $\epsilon_0^{(1,2)}$ for the full evolution super-operator is a formidable task. However, when the regime $|\mathcal{J}_{ij}| \ll 1$, $V \gg T$ is considered, the unitary dynamics of the magnetic impurity is much faster than its relaxation dynamics. This allows us to use the rotating wave approximation (RWA) [S2], and calculate $\epsilon_0^{(1)}$ and $\epsilon_0^{(2)}$ considering the evolution of only the diagonal components of the density matrix (in the eigenbasis of $H_{\rm e-i}^{\rm mf}$). Within the RWA we obtain from Eq. (S7)

$$\frac{d}{dt} \langle m | \rho_S^{(\lambda)}(t) | m \rangle = \sum_{m=-S}^{S} \mathcal{L}_{mm'}^{(\lambda)} \langle m' | \rho_S^{(\lambda)}(t) | m' \rangle. \tag{S16}$$

Here $|m\rangle$ with $m = S, S-1, ..., -S$ are the eigenstates of $H_{\rm e-i}^{\rm mf}$. Since the exchange matrix $\mathcal{J}_{ij}$ is assumed to be lower triangular (see the discussion of the model in the main text), the mean-field interaction is directed along $z$ axis, $H_{\rm e-i}^{\rm mf} = \mathcal{J}_{zz} S_z V/2$. Therefore, states $|m\rangle$ are eigenstates of $S_z$: $S_z |m\rangle = m |m\rangle$. The matrix $\mathcal{L}_{mm'}^{(\lambda)}$ has the following non-zero elements

$$\begin{cases} \mathcal{L}_{m\pm 1, m}^{(\lambda)} = e^{-i\lambda} \eta_\pm^{(0)} (S(S+1) - m(m \pm 1))/4, \\ \mathcal{L}_{m,m}^{(\lambda)} = \left(e^{-i\lambda} - 1\right) \eta_{zz}^{(0)} m^2 - e^{i\lambda} \left(\mathcal{L}_{m-1,m}^{(\lambda)} + \mathcal{L}_{m+1,m}^{(\lambda)}\right). \end{cases} \tag{S17}$$

To find the coefficients $\epsilon_0^{(1)}$ and $\epsilon_0^{(2)}$ we calculate the shift of zero eigenvalue of a non-Hermitian matrix $\mathcal{L}^{(0)}$ mediated by the perturbation $\Delta \mathcal{L}^{(\lambda)} = \mathcal{L}^{(\lambda)} - \mathcal{L}^{(0)}$ up to the second order in $\lambda$.

### A. First order in $\lambda$: average backscattering current

We start by calculating the coefficient $\epsilon_0^{(1)}$ using the first order non-degenerate perturbation theory in $\lambda$. First we find the right and left eigenvectors of $\mathcal{L}^{(0)}$ with the eigenvalue $\epsilon_0(\lambda = 0) = 0$. The right eigenvector corresponds to the stationary state solution for the density matrix at $\lambda = 0$. Direct check shows that $\mathcal{L}^{(0)} |0\rangle\!\rangle = 0$ for

$$|0\rangle\!\rangle = \frac{1}{\mathcal{Z}} \left(\vartheta^S, \vartheta^{S-1}, ..., \vartheta^{-S}\right)^T, \qquad \vartheta = \frac{1+p}{1-p}, \qquad \mathcal{Z} = \sum_{m=-S}^{S} \vartheta^m. \tag{S18}$$

The left eigenvector $\langle\!\langle \tilde{0} |$, obeying $\langle\!\langle \tilde{0} | \mathcal{L}^{(0)} = 0$, is given by

$$\langle\!\langle \tilde{0} | = (1, 1, ..., 1) \tag{S19}$$

as dictated by the conservation of the trace of the density matrix at $\lambda = 0$. In formulae above we introduced double bra and ket notations to distinguish eigenvectors of $\mathcal{L}^{(\lambda)}$ from usual quantum-mechanical states. The vectors (S18) and (S19) are normalized so that $\langle\!\langle \tilde{0} | 0 \rangle\!\rangle = 1$.

As a next step we write down explicitly the perturbation matrix:

$$\Delta \mathcal{L}_{mm'}^{(\lambda)} = (e^{-i\lambda} - 1)\left[\eta_{zz}m^2 \delta_{mm'} + (1-\delta_{mm'})\mathcal{L}_{mm'}^{(0)}\right] = (e^{-i\lambda}-1)\mathcal{L}_{mm'}^{(0)} + (e^{-i\lambda}-1)\delta_{mm'}\left[\eta_{zz}m^2 - \mathcal{L}_{mm}^{(0)}\right]. \quad \text{(S20)}$$

Because $\langle\!\langle \tilde{0}|\mathcal{L}^{(0)} = 0$ and $\mathcal{L}^{(0)}|0\rangle\!\rangle = 0$, the part of $\Delta \mathcal{L}_{mm'}^{(\lambda)}$ that is proportional to $\mathcal{L}_{mm'}^{(0)}$ does not generate a shift of the eigenvalue $\epsilon_0$ in the first two orders of perturbation theory in $\lambda$. After some straightforward algebra we find that the part of the perturbation that affects zero eigenvalue in the first two orders of the perturbation theory is given by the following *diagonal* operator:

$$\Delta \widetilde{\mathcal{L}}^{(\lambda)} = (e^{-i\lambda}-1)\frac{\pi gV}{4}R(S_z), \qquad R(S_z) = (2-3q)S_z^2 - pqS_z + qS(S+1). \quad \text{(S21)}$$

Then we obtain

$$\epsilon_0^{(1)} = \left.\langle\!\langle \tilde{0}|\partial_\lambda \Delta\widetilde{\mathcal{L}}^{(\lambda)}|0\rangle\!\rangle\right|_{\lambda=0} = -i\frac{\pi gV}{4}\langle R(S_z)\rangle, \qquad \langle R(S_z)\rangle = \sum_{m=-S}^{S}\frac{\vartheta^m}{\mathcal{Z}}R(m). \quad \text{(S22)}$$

Equation (S22) results in Eq. (8) of the main text, which determines the average backscattering current.

### B. Second order in $\lambda$: zero-frequency noise and backscattering Fano factor

The second order correction to the zero eigenvalue of $\mathcal{L}^{(0)}$ consists of two contributions, $\epsilon_0^{(2)} = \epsilon_0^{(2,1)} + \epsilon_0^{(2,2)}$. The first contribution, $\epsilon_0^{(2,1)}$, is obtained by expanding the perturbation matrix $\Delta\widetilde{\mathcal{L}}^{(\lambda)}$ to the second order in $\lambda$ and then applying first order perturbation theory. The resulting expression is

$$\epsilon_0^{(2,1)} = \left.\langle\!\langle \tilde{0}|\partial_\lambda^2 \Delta\widetilde{\mathcal{L}}^{(\lambda)}|0\rangle\!\rangle\right|_{\lambda=0} = -\frac{\pi gV}{4}\langle R(S_z)\rangle. \quad \text{(S23)}$$

Finding $\epsilon_0^{(2,2)}$ requires a more cumbersome procedure of second order perturbation theory. The first step is to find orthogonal complements to the vectors $\langle\!\langle \tilde{0}|$ and $|0\rangle\!\rangle$. We consider the following sets of vectors:

$$|n\rangle\!\rangle = \frac{1}{\mathcal{Z}}(\underbrace{\vartheta^S, \vartheta^{S-1}, ..., \vartheta^{S-n+1}}_{n\text{ components}}, 0, 0, ..., 0)^T - \sum_{m=S-n+1}^{S}\frac{\vartheta^m}{\mathcal{Z}}|0\rangle\!\rangle, \qquad n=1,...,2S, \quad \text{(S24)}$$

$$\langle\!\langle \tilde{r}| = (\underbrace{1,1,1,...,1,1}_{r\text{ components}}, 0, 0, ..., 0) - \sum_{m=S-r+1}^{S}\frac{\vartheta^m}{\mathcal{Z}}\langle\!\langle \tilde{0}|, \qquad r=1,...,2S. \quad \text{(S25)}$$

Direct check shows that sets $|n\rangle\!\rangle$ and $\langle\!\langle \tilde{r}|$ are orthogonal to $\langle\!\langle \tilde{0}|$ and $|0\rangle\!\rangle$, respectively, i.e., $\langle\!\langle \tilde{0}|n\rangle\!\rangle = 0$ and $\langle\!\langle \tilde{r}|0\rangle\!\rangle = 0$ for $n,r=1,...,2S$. The vector sets introduced here are particularly appealing because they diagonalize $\mathcal{L}^{(0)}$:

$$\langle\!\langle \tilde{r}|\mathcal{L}^{(0)}|n\rangle\!\rangle = -\delta_{rn}\frac{\pi gV}{2}q(1-p)\frac{S(S+1)-(S+1-n)(S-n)}{4}\frac{\vartheta^{S+1-n}}{\mathcal{Z}}, \qquad r,n=1,...,2S. \quad \text{(S26)}$$

Note, however, that $\langle\!\langle \tilde{r}|n\rangle\!\rangle \neq 0$. Overall, we find

$$\epsilon_0^{(2,2)} = -2\sum_{n=1}^{2S}\frac{\left.\langle\!\langle \tilde{0}|\partial_\lambda \Delta\widetilde{\mathcal{L}}^{(\lambda)}|n\rangle\!\rangle\right|_{\lambda=0}\left.\langle\!\langle \tilde{n}|\partial_\lambda \Delta\widetilde{\mathcal{L}}^{(\lambda)}|0\rangle\!\rangle\right|_{\lambda=0}}{\langle\!\langle \tilde{n}|\mathcal{L}^{(0)}|n\rangle\!\rangle}. \quad \text{(S27)}$$

After straightforward algebraic manipulations we obtain

$$\epsilon_0^{(2,2)} = -\frac{\pi gV}{q(1-p)}\sum_{n=1}^{2S}\frac{\langle\!\langle \tilde{0}|R(S_z)|n\rangle\!\rangle\langle\!\langle \tilde{n}|R(S_z)|0\rangle\!\rangle}{n(2S+1-n)}\frac{\mathcal{Z}}{\vartheta^{S+1-n}}. \quad \text{(S28)}$$

The explicit structure of vectors $\langle\!\langle \tilde{r}|$ and $|n\rangle\!\rangle$ allows us to rewrite this expression in terms of the averages of $R(S_z)$. To this end we note that

$$\langle\!\langle \tilde{0}|R(S_z)|n\rangle\!\rangle = \langle\!\langle \tilde{n}|R(S_z)|0\rangle\!\rangle = \langle P_n\left[R(S_z) - \langle R(S_z)\rangle\right]\rangle, \qquad P_n = \sum_{m=S-n+1}^{S}|m\rangle\langle m|. \quad \text{(S29)}$$

Therefore, the expression for $\epsilon_0^{(2,2)}$ simplifies to

$$\epsilon_0^{(2,2)} = -\frac{\pi g V}{q(1-p)} \sum_{n=1}^{2S} \frac{\langle P_n\left[R(S_z) - \langle R(S_z)\rangle\right]\rangle^2}{n(2S+1-n)} \frac{\mathcal{Z}}{\vartheta^{S+1-n}}. \tag{S30}$$

Combining $\epsilon_0^{(2,1)}$ and $\epsilon_0^{(2,2)}$ and introducing $\mu_n = \vartheta^{S+1-n}/\mathcal{Z}$, we obtain the expression for the zero-frequency noise:

$$\mathcal{S}_{\rm bs} = \frac{\pi g V}{4} \langle R(S_z)\rangle \left( 1 + \frac{4}{q(1-p)} \sum_{n=1}^{2S} \frac{\langle P_n\left[R(S_z) - \langle R(S_z)\rangle\right]\rangle^2}{n(2S+1-n)\langle R(S_z)\rangle \mu_n} \right). \tag{S31}$$

Eqs. (S22) and (S31) lead to Eq. (9) in the main text for the backscattering Fano factor $F_{\rm bs}$. We stress that, despite the $(1-p)$ factor in the denominator of Eq. (S31), the result is actually finite for $p \to 1$, since then the impurity becomes fully polarized and $\langle P_n\left[R(S_z) - \langle R(S_z)\rangle\right]\rangle \to 0$.

Finally, let us briefly discuss the effects of electron-electron interactions at the helical edge on the average backscattering current and the zero-frequency noise. As was shown in Sec. S.I, the matrix structure of the correlation function $\Pi_V^{(\lambda)}$ is insensitive to the presence of interactions in the shot noise limit $V \gg T$ (compare Eqs. (S10) and (S13)): nonzero interaction just results in a voltage-dependent multiplicative factor. This feature allows us to modify the expressions for $I_{\rm bs}$ and $\mathcal{S}_{\rm bs}$ to account for $K \neq 1$:

$$I_{\rm bs} = \frac{\pi g V \langle R(S_z)\rangle}{4\Gamma(2K)} \left(\frac{aV}{u}\right)^{2K-2}, \quad \mathcal{S}_{\rm bs} = \frac{\pi g V \langle R(S_z)\rangle}{4\Gamma(2K)} \left(\frac{aV}{u}\right)^{2K-2} \left( 1 + \frac{4}{q(1-p)} \sum_{n=1}^{2S} \frac{\langle P_n\left[R(S_z) - \langle R(S_z)\rangle\right]\rangle^2}{n(2S+1-n)\langle R(S_z)\rangle \mu_n} \right). \tag{S32}$$

Strikingly, the ratio between $\mathcal{S}_{\rm bs}$ and $I_{\rm bs}$ – that is, the backscattering Fano factor $F_{\rm bs}$ – does not depend on $K$ at all, and thus is insensitive to the electron-electron interaction at the helical edge.

---



\end{document}